\documentclass[11pt]{article}
\usepackage{amsmath,amssymb}

\usepackage{epsfig}  
\usepackage{graphicx}               
\usepackage{url}
\usepackage{hyperref}
\usepackage{bm}
\usepackage{slashed}
\usepackage{enumitem}
\usepackage{cite}
\usepackage[dvipsnames]{xcolor}

\usepackage{pstricks}
\usepackage{color}

\newcommand{\nc}{\newcommand}  

\nc{\beq}{\begin{equation}}  
\nc{\eeq}{\end{equation}}  
\nc{\beqa}{\begin{eqnarray}}  
\nc{\eeqa}{\end{eqnarray}}  
\nc{\bit}{\begin{itemize}}  
\nc{\eit}{\end{itemize}}

\newcommand{\ts}{\thinspace}

\title{  
\vspace*{-2.3cm}
\begin{flushright}
\normalsize{
  }
\end{flushright}
\vspace{1.5cm}
\Large  
\textbf{Search for Heavy Stops with Merged Top-Jets
}\vspace*{1.0cm}   
}

\author{Yang Bai, Joshua Berger, James Osborne and Ben A. Stefanek
\vspace{5mm}
\\
\normalsize\emph{Department of Physics, University of Wisconsin-Madison, Madison, WI 53706, USA}  
}
\date{}

\begin{document}  
\setcounter{page}{0}  
\maketitle  

\vspace*{1cm}  
\begin{abstract} 
We study an interesting region of phase space at the LHC for pair-produced stops decaying into hadronic top quarks and light neutralinos. After imposing a sizeable cut on the missing transverse energy, which is the key variable for reducing backgrounds, we have found that the two hadronic tops are likely to merge into a single fat jet. We develop a jet-substructure-based strategy to tag the two merged top-jets and utilize the $M_{T2}$ variable to further reduce the backgrounds. We obtain about a 50\% increase to the ratio of the signal over background and a mild increase on the signal discovery significance, based on a signal with a 1.2 TeV stop and a 100 GeV neutralino, for the 13 TeV LHC with 100 fb$^{-1}$. The general event kinematics could also occur and be explored for other new physics signatures with large missing transverse energy. 
\end{abstract}  
  
\thispagestyle{empty}  
\newpage  
  
\setcounter{page}{1}  
  
\baselineskip18pt   

\vspace{-2cm}

\section{Introduction}
\label{sec:intro}
While the discovery of the Higgs boson \cite{CMS:2012nga,ATLAS:2012oga} was a major success for the Standard Model (SM), open theoretical questions remain. If the Higgs boson is an elementary scalar field, the stability of the electroweak scale against radiative corrections is not understood, since fields of this type receive quadratically divergent corrections to their mass-squared. Of the proposed solutions to this well known hierarchy problem, one of the best motivated is supersymmetry (SUSY), where every SM fermion(boson) is complemented with a superpartner boson(fermion). Elementary scalar masses are protected because a cancellation occurs between the quadratically divergent correction coming from loops with SM particles and their corresponding superpartners enforced by the enlarged symmetry of the model. Nevertheless, SUSY must be broken in order to be consistent with non-observation of superpartners to date. This breaking lifts the superpartner masses to a relatively high scale set by the SUSY breaking sector and messenger mass~\cite{Martin:1997ns,Chung:2003fi}. 

As the superpartners become more massive, the expected mass of elementary scalars goes up and fine-tuning is likely required to explain light scalars. Since the Higgs boson receives the largest correction to its mass from a loop involving top quarks, the most important SUSY particle for protecting the Higgs boson mass is the superpartner of the top quark--the top squark or stop~\cite{Dimopoulos:1995mi,Cohen:1996vb}. In the Minimal Supersymmetric Standard Model (MSSM), the electroweak scale is dominantly set by stop masses and mixing, in addition to the tree level mass scales in the superpotential and soft SUSY breaking potential. Thus, the natural expectation is that the lightest third generation squarks should be observable at the Large Hadron Collider (LHC), since the mass scales that set the electroweak scale in the MSSM must not be too far away from the observed electroweak scale in order to avoid large amounts of fine tuning. While a considerable amount of natural MSSM parameter space has been excluded by the most recent results from the LHC~\cite{Cahill-Rowley:2014twa,Baer:2014ica}, viable regions remain, such as the compressed region \cite{Han:2012fw,Delgado:2012eu,Dutta:2013gga,Cho:2014yma,Ferretti:2015dea,An:2015uwa,Belanger:2015vwa,Kobakhidze:2015scd,Cheng:2016npb}.  These regions are likely difficult to achieve in a top down SUSY breaking model.

In this paper, we accept some fine tuning and focus on a new method to search for pair-produced stops with a mass near 1 TeV which decay to top quarks and light neutralinos, a scenario which has not yet been excluded by the LHC Run 2. We assume the neutralino to be the lightest supersymmetric particle (LSP) with a mass around 100 GeV, which also makes it a possible thermal relic weakly interacting massive particle (WIMP) dark matter candidate. The signal that our new method will be applicable to is $\bar{t}t+ \slashed{E}_{\rm T}$ (see Ref.~\cite{Plehn:2010st,Bi:2011ha,Plehn:2012pr,Alves:2012ft,Kaplan:2012gd,Buckley:2013lpa} for earlier studies of two hadronic tops), where the tops decay hadronically and the missing transverse energy $\slashed{E}_{\rm T}$ comes from the two neutralinos which leave the detector. Currently, for searches with two hadronic tops, the ATLAS collaboration has imposed a constraint which requires the stop mass to be above 820 GeV with 13.3 $\rm fb^{-1}$ of integrated luminosity, assuming 100\% decay branching for the channel $\widetilde{t} \rightarrow t\,\widetilde{\chi}_{0}$~\cite{ATLAS-CONF-2016-077} and a light neutralino mass below around 200 GeV. Similarly, the CMS collaboration has obtained a limit of 860 GeV with 12.9 fb$^{-1}$ data~\cite{CMS-PAS-SUS-16-029}. As the $\slashed{E}_{\rm T}$ cut is increased to optimize for heavier stops, it becomes increasingly likely that the two tops in the decay have a small angular separation and that the signals of their decay products overlap in the detectors. 

We develop a new boosted top tagging procedure that recovers merged top jets, which we call ``Merged Top Tagger.''  Our starting point is the well-known HEPTopTagger algorithm \cite{Plehn:2010st,Kasieczka:2015jma}.  Rather than simply trying to find a single combination of subjets in a fat jet that looks like a top, the algorithm searches through various combinatoric possibilities in an attempt to find a total of two top-like groups of subjets.  Further sensitivity is gained by using the new information of both tops' kinematics to construct and cut on the $M_{T2}$ variable \cite{Lester:1999tx}. Based on simulations we have performed, our strategy shows a 50\% improvement in $S/B$ and slight improvement in discovery sensitivity for $1.2~{\rm TeV}$ stops decaying to tops and $100~{\rm GeV}$ detector-stable neutralinos.  

The remainder of this paper is structured as follows.  In Section~\ref{sec:motiv}, we motivate the need for a new analysis strategy to deal with merged hadronic tops.  We present our Merged Top Tagger algorithm in Section~\ref{sec:algo}.  We then study the increased sensitivity to search for heavy stops in Section~\ref{sec:results}.  Finally, we discuss and conclude in Section~\ref{sec:conc}. The simulation detail is presented in Appendix~\ref{app:simulation}.

\section{Motivation for Merged Top-Jets}
\label{sec:motiv}

\begin{figure}[ht!]
  \centering
  \includegraphics[scale=0.6]{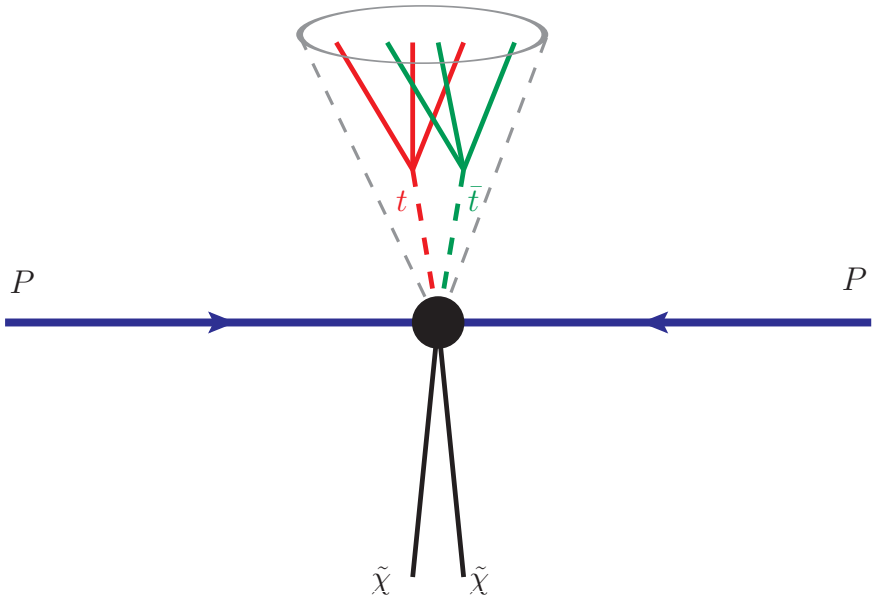}
  \caption{The schematic plot of the region of phase space for signal events with a large cut on $\slashed{E}_{\rm T}$. To provide large $\slashed{E}_{\rm T}$, the neutralino momenta must approximately align. This also results in alignment of the tops, which recoil off the sum of the neutralino momenta. When performing jet-substructure analysis with a large value of the jet clustering parameter $R$, the six partons from the two hadronic tops are likely to be included in a single fat jet.}
  \label{fig:schematic}
\end{figure}

For the region of parameter space with $m_{\widetilde{t}}-m_{\widetilde{\chi}_{0}} \gg m_t$, the top quark is boosted and the three partons from the top quark decay are collimated. In similar scenarios with boosted top quarks, both collaborations at the LHC have opted to reconstruct events using jet-substructure techniques such as top-tagging. Assuming conservation of R-parity, the two neutralinos in the final state will be stable and contribute significantly to the total missing transverse energy, $\slashed{E}_{\rm T}$. Because of this fact, imposing a large $\slashed{E}_{\rm T}$ cut is a very efficient way to increase the signal over background ratio. However, we find that imposing a large  $\slashed{E}_{\rm T}$ cut isolates the sub-region of phase space of the signal where the neutralinos are approximately aligned to provide large $\slashed{E}_{\rm T}$, resulting in the tops (which recoil off the neutralino momenta) also being approximately aligned. Because the pair-produced stops are not very relativistic, the top and neutralino pairs end up approximately back to back. As a result, one should anticipate that a significant fraction of signal events will have geometric overlapping between the six partons from two hadronic tops, a scenario we call {\it merged top-jets}. The schematic picture of the region of phase space for signal events under discussion is shown in Fig.~\ref{fig:schematic}. The subject of this paper will be a new search strategy for the merged top-jets case, with the goal of improving the discovery potential of stops at the LHC. 

\begin{figure}[h!]
  \centering
  \includegraphics[scale=0.62]{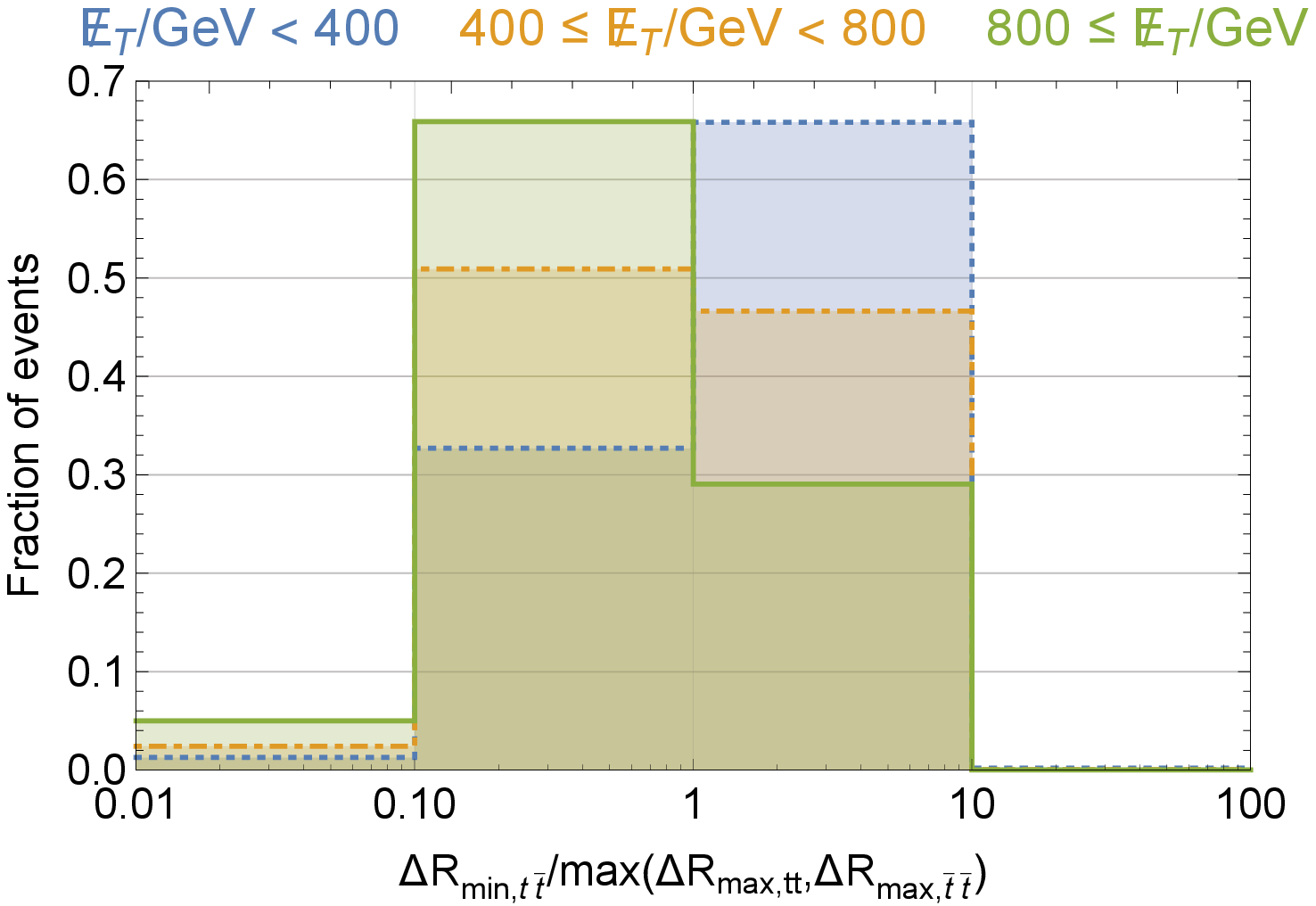} \hspace{5mm}
  \includegraphics[scale=0.48]{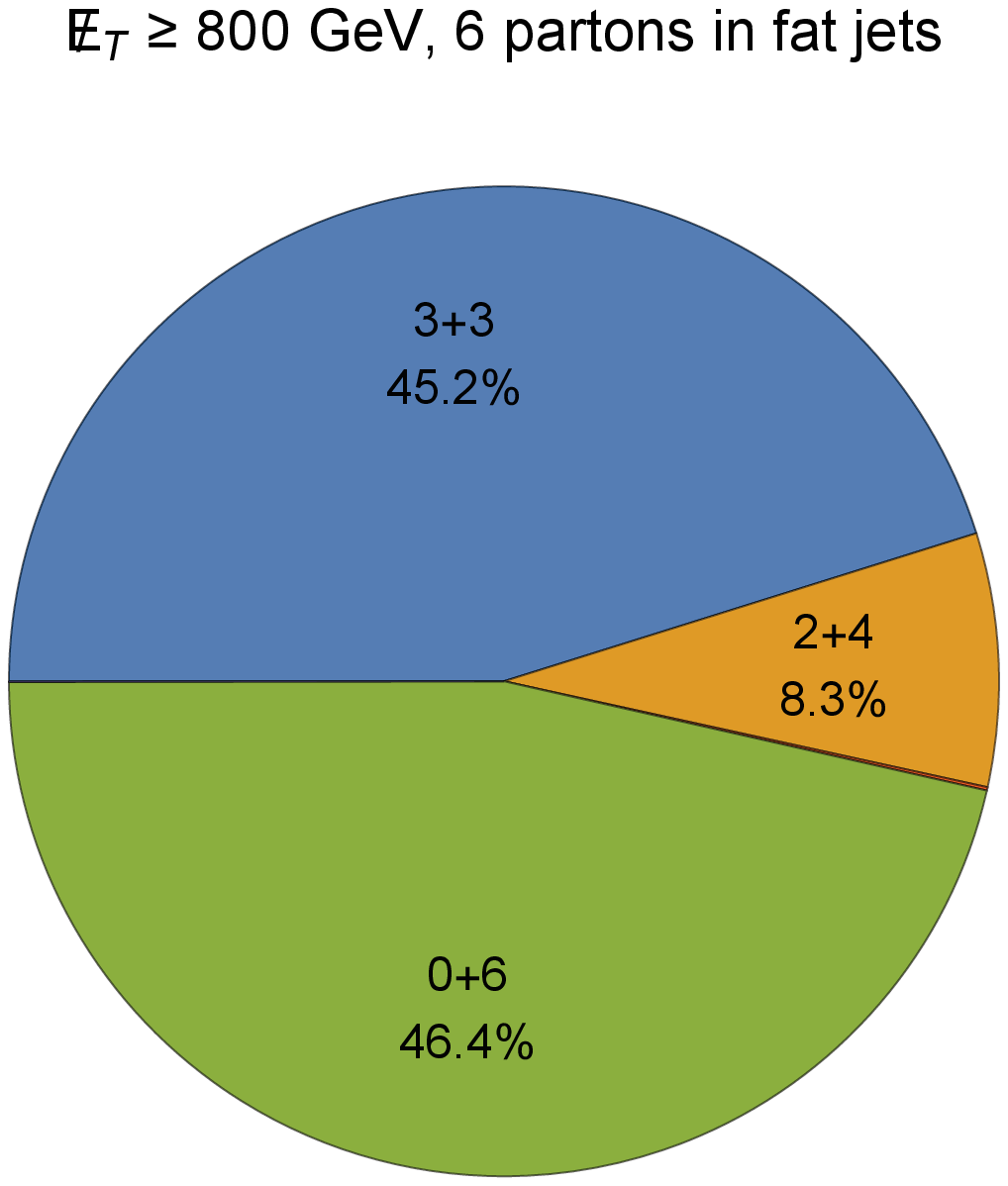}
  \caption{Left panel: The fraction of signal events as a function of a variable (see text for definition) which measures the overlapping of partons from the two tops: a smaller value means a larger overlap. Right panel: The distribution of partons in a $R=1.5$ Cambridge/Aachen fat jet and $p_T > 200$~GeV. Not shown are the negligible fractions of events with 1+5 and 1+2+3.}
  \label{fig:parton_splitting}
\end{figure}

To quantitatively understand the fraction of signal events with two merged tops, we first study the signal events at the parton level. Throughout this paper, we will assume  benchmark masses of $m_{\widetilde{t}} = 1.2$~TeV and $m_{\widetilde{\chi}}=100$~GeV, and $\mbox{Br}(\widetilde{t} \rightarrow t + \widetilde{\chi}) = 100\%$. The lightest stop mass is chosen to be close to the reach of the LHC Run 2 with 100 fb$^{-1}$. In our simulation, we have the stop to be mainly right-handed, although our later kinematic analysis is insensitive to this choice (for the detailed analysis for left-handed and right-handed stops, see Ref.~\cite{Low:2013aza}). In the left panel of Fig.~\ref{fig:parton_splitting}, we define a measure to demonstrate the alignment of the two tops as the cut on $\slashed{E}_{\rm T}$ is increased.
To show this, we define $\Delta R_{{\rm min},t\bar{t}}$ as the minimum geometrical separation among all pairs with one parton from the top and one parton from the anti-top. Then we define ${\rm max} (\Delta R_{{\rm max},tt},\Delta R_{{\rm max}, \bar{t}\bar{t}})$ as the maximum separation among all pairs with two partons belonging only to the top or anti-top. The ratio of these two quantities, $\Delta R_{{\rm min},t\bar{t}} \ts / \ts {\rm max} (\Delta R_{{\rm max},tt},\Delta R_{{\rm max},\bar{t}\bar{t}})$, will be small when the smallest distance between one parton from the top and one parton from the anti-top is smaller than the largest distance between partons in the same top or anti-top, which is the aligned or merged scenario. As a result, we expect the event distribution of this measure to shift to smaller values as the cut on the missing transverse energy is increased, and this expectation is confirmed in the left panel of Fig.~\ref{fig:parton_splitting}. 

When two partons from the top and anti-top are geometrically close to each other, the standard jet-substructure analysis to tag boosted top-jets becomes problematic. This is because the large fat-jet analysis employed frequently fails to isolate the two tops in separate fat jets. To demonstrate this issue at parton level, we apply the Cambridge/Aachen clustering algorithm~\cite{Dokshitzer:1997in} with $R=1.5$ to the signal events and count the number of partons belonging to the top or anti-top contained in each fat jet. In the right panel of Fig.~\ref{fig:parton_splitting}, one  can see that the fully merged 0+6 case (all partons in one fat jet) and the partially merged 2+4 case (2 partons in one fat jet, 4 partons in another) account for 54.7\% of the events, with the separated 3+3 case accounting for 45.2\% of the events. The 1+5 and 1+2+3 cases are extremely sub-dominant due to the fat jet definition including a transverse momentum cut of $p_{T} >200$~GeV. 

It is now clear that the standard jet-substructure analysis is not optimal for reconstructing our signal events after a large $\slashed{E}_{\rm T}$ cut, since standard top taggers (which assume well isolated tops by only searching for one top tag per fat jet) frequently fail to tag two tops in the fully or partially merged cases which constitute a majority of our signal events after a large $\slashed{E}_{\rm T}$ cut. To fully recover the signal kinematics and define additional transverse-mass variables like $M_{T2}$, it is important to resolve and tag the two top-jets with high efficiency. In next section of this paper, we develop a search strategy based on the HEPTopTagger~\cite{Plehn:2010st,Kasieczka:2015jma} algorithm and check the improvement of signal over background after the particle-level simulations.

\section{The Merged Top Tagger Algorithm}
\label{sec:algo}
Having established our motivation for tagging two merged top-jets, we now develop a jet-substructure based algorithm to reconstruct both boosted hadronic tops for the signal and background. Our algorithm for tagging two merged top-jets is based on the existing top-tagging algorithm of HEPTopTagger in Ref.~\cite{Plehn:2010st}, although it can also be implemented using other top-tagging methods such as  N-subjettiness~\cite{Thaler:2011gf}. Before discussing our algorithm, we briefly summarize the features of HEPTopTagger. By default, HEPTopTagger takes a single fat jet as an input, on which it will perform a mass drop~\cite{Butterworth:2008iy} operation to obtain a list of the relevant hard substructure. It then iterates through all pairings of three hard subjets and filters~\cite{Butterworth:2008iy} each triplet in order to define the triplet mass. It then keeps {\it only} the triplet which has a filtered mass closest to the top mass $m_{t}$. This single remaining triplet is then required to pass additional mass constraints which ensures the triplet does not behave like a QCD event and that it satisfies some top and $W$ gauge boson mass constraints. 

For our signal events, after a large $\slashed{E}_{\rm T}$ cut, the two top-jets have a significant probability of being merged, and the original HEPTopTagger algorithm becomes less efficient. This is because the original HEPTopTagger algorithm can find at most one top per fat jet, even in the case where the input fat jet contains two tops. Furthermore, even in the one top per fat jet case, the original algorithm will fail when the triplet with the best filtered mass fails the mass criteria, even though there may exist other triplets with filtered masses still not that far away from the top quark mass, but a better chance to pass all mass criteria. To address those issues, we introduce the following ``Merged Top Tagger" algorithm with an aim to tag two merged top-jets: 
%

\begin{enumerate}
\item Using the Cambridge/Aachen algorithm with $R = 1.5$, identify initial fat jets with $p_T(J) >$~200 GeV.
\item For the leading $p_{T}$ fat jet, find all hard subjets using a mass drop criterion: when undoing the last clustering of the jet $j$, into two subjets $j_1$, $j_2$ with $m_{j_1} > m_{j_2}$, we require $m_{j_1} < 0.8 \ts\ts m_j$ to keep $j_1$ and $j_2$. Otherwise, we keep only $j_1$. For each subjet $j_i$, we further de-cluster it until its jet mass is 30 GeV or below. 
\item Iterate through all pairings of three hard subjets (HSJ triplets) found in Step 2: first, filter them with resolution $R_{\rm filter} = {\rm min}(0.3, \Delta R_{jk}/2)$. Next, use the five hardest filtered constituents and calculate the triplet mass $m_{\rm filtered}$ (for less than five filtered constituents use all of them). Keep all HSJ triplets which satisfy $m_{t_{\rm min}} < m_{\rm filtered} < m_{t_{\rm max}}$ (Default: $m_{t_{\rm min}} =$ 140 GeV and $m_{t_{\rm max}} =$ 250 GeV). 
\item For each HSJ triplet saved from Step 3, re-cluster the five filtered constituents into exactly three subjets $j_1$, $j_2$, $j_3$, ordered by $p_T$. If the masses
($m_{12}$, $m_{13}$, $m_{23}$) satisfy one of the following three criteria, save the filtered HSJ triplet as a top candidate:
\begin{equation*}
0.2 < {\rm arctan} \ts \left(\frac{m_{13}}{m_{12}}\right) < 1.3 \hspace{7.5mm} {\rm and} \hspace{7.5mm} R_{\rm min} < \frac{m_{23}}{m_{123}} < R_{\rm max}  \,,
\end{equation*}
\begin{equation*}
R^{2}_{\rm min} \left[ 1+ \left( \frac{m_{13}}{m_{12}}\right)^{2} \right] < 1 -  \left( \frac{m_{23}}{m_{123}}\right)^{2} < R^{2}_{\rm max} \left[ 1+ \left( \frac{m_{13}}{m_{12}}\right)^{2} \right]  \hspace{7.5mm} {\rm and} \hspace{7.5mm} \frac{m_{23}}{m_{123}} > 0.35 \,,
\end{equation*}
\begin{equation*}
R^{2}_{\rm min} \left[ 1+ \left( \frac{m_{12}}{m_{13}}\right)^{2} \right] < 1 -  \left( \frac{m_{23}}{m_{123}}\right)^{2} < R^{2}_{\rm max} \left[ 1+ \left( \frac{m_{12}}{m_{13}}\right)^{2} \right]  \hspace{7.5mm} {\rm and} \hspace{7.5mm} \frac{m_{23}}{m_{123}} > 0.35 \,.
\vspace{2mm}
\end{equation*}
with $R_{\rm min} = (1-f_{W}) \times M_W /m_t$ and $R_{\rm max} = (1+f_{W}) \times M_W /m_t$ (we will choose the value $f_{W} = 0.5$~\cite{CMS:2014fya} to increase signal efficiencies). The number 0.35 is chosen to help remove QCD events. The above selection criteria are identical to HEPTopTagger~\cite{Plehn:2010st}. 
\item If there is more than one top candidate from Step 4, check all two pairings of top candidates and keep any for which their HSJ triplets share no subjets (unique pairs). If only one unique pair is found, return it as two tagged tops. If more than one unique pair is found, return the pair of top candidates which minimizes the quantity $|m_{J_{1,{\rm filtered}}} - m_t| + |m_{J_{2,{\rm filtered}}} - m_t|$ as tagged tops. Otherwise, continue to Step 6.
\end{enumerate}
Up to this point, the modified HEPTopTagger algorithm has only been extended to deal with the fully merged case when two tops are contained within a single fat jet. To deal with the partially merged case when some partons from the first top are clustered into the fat jet of the second top, we also introduce the following steps to capture particles in the vicinity of the leading fat jet, which we accomplish by removing the particles belong to the leading top candidate(s) and reclustering the event. Specifically, we extend the algorithm in the following way:
\begin{enumerate}[resume]
\item If at least one top candidate exists from Step 4, remove the particles associated with one top candidate from the final state particles. The remaining particles are then reclustered using the same Cambridge/Aachen algorithm with $R=1.5$. For the leading $p_{T}$ fat jet found after this reclustering (if any), repeat the above Steps 2-4 to identify more top candidates. Of all top candidates found in the reclustered fat jet, keep the one which has a filtered mass closest to $m_{t}$ (if any), and pair it with the top candidate before the new reclustering procedure. For the cases with more than one top candidate from Step 4, perform this procedure on all of them, and return the pair which minimizes the quantity $|m_{J_{1,{\rm filtered}}} - m_t| + |m_{J_{2,{\rm filtered}}} - m_t|$ as tagged tops. If no top candidate pairs are found, continue to Step 7.
\item If the two leading $p_{T}$ fat jets from the initial event clustering have not both already been analyzed, repeat Steps 2-6 on the next leading initial fat jet by $p_{T}$, if it exists. If no initial fat jets remain or if the two leading $p_{T}$ initial fat jets have already been analyzed, then the algorithm has failed to tag two tops. Of all the top candidates from the leading $p_{T}$ initial fat jet, return the one that minimizes $|m_{J_{{\rm filtered}}} - m_t| $ as a single tagged top. If there are no top candidates in the leading $p_{T}$ initial fat jet, use the next leading $p_{T}$ initial fat jet. If there are no top candidates in any initial fat jet, the algorithm has failed to tag a single top.
\end{enumerate}
%

Not only does our modified algorithm allow two tops to be tagged in the fully and partially merged cases, it also finds the combination of hard subjets such that both tops are as close to the true top mass as possible. To quantify the improvement our algorithm offers for tagging merged top-jets, we define the following efficiency parameter 
\begin{equation}
\mathcal{E}_{N_{t}\geq 2} \ts (\slashed{E}_{\rm T} \ts {\rm range}) \equiv \frac{\text{Events with $ \geq 2$ top tags in $\slashed{E}_{\rm T} \ts {\rm range}$}}{\text{All events in $\slashed{E}_{\rm T} \ts {\rm range}$}} \ts ,
\end{equation} 
which we can use to gauge the relative performance of our modified algorithm versus the original HEPTopTagger algorithm over different ranges of missing transverse energy. According to our hypothesis, a larger cut on $\slashed{E}_{\rm T}$ should increase the number of merged top signal events, pushing more signal events into the region where the original algorithm cannot tag two tops. As a result, we expect the ability of the modified algorithm to tag two tops to increase relative to the original algorithm for higher missing energy windows, an expectation which is confirmed in Table~\ref{tab:signal_eff}. 
\begin{table}[h!]
\renewcommand{\arraystretch}{1.5}
    \resizebox{\linewidth}{!}{
    \begin{tabular}{| c | c | c || c | c || c | c |}
    \hline
   & \multicolumn{2}{|c||}{Signal} & \multicolumn{2}{|c||}{$\overline{t}t+{\rm jets}$} &  \multicolumn{2}{|c|}{$\overline{t}t + Z$}\\
    \hline \hline 
    Algorithm &   $\slashed{E}_{\rm T} > 400$~GeV & $\slashed{E}_{\rm T} > 800$~GeV  & $\slashed{E}_{\rm T} > 400$~GeV & $\slashed{E}_{\rm T} > 800$~GeV & $\slashed{E}_{\rm T} > 400$~GeV & $\slashed{E}_{\rm T} > 800$~GeV\\
    \hline
    Original & 18\% & 14\% & 6.0\% & 7.1\% & 8.4\% & 11\% \\
    \hline
    Modified & 48\% & 55\% & 18\% & 26\% & 23\% & 41\% \\
    \hline \hline
    Ratio & 2.6 & 3.9 & 3.0 & 3.6 & 2.8 & 3.6  \\
    \hline
  \end{tabular}
  }
  \caption{The efficiencies, $\mathcal{E}_{N_{t}\geq 2}$, to tag two top-jets using the original HEPTopTagger and our Merged Top Tagger. The signal events have $m_{\widetilde{t}} = 1.2$~TeV and $m_{\widetilde{\chi}}=100$~GeV. Only the dominant backgrounds are shown here for comparison.}
  \label{tab:signal_eff}
\end{table}
Also shown in Table~\ref{tab:signal_eff} are the two top-tagging efficiencies for the two leading backgrounds, $\overline{t}t+ {\rm jets}$ (one hadronic top and one leptonic top) and $\overline{t}t + Z$ (two hadronic tops, $Z$ decays to two neutrinos). Having discussed the tagging efficiency for our new algorithm, we turn now to estimate the improvement for the signal over background ratio and the signal discovery significance. 

\section{Estimation of Discovery Significance}
\label{sec:results}
In this section, we estimate the discovery sensitivity of stops in the fully-hadronic channel at the LHC Run 2 with 100~fb$^{-1}$.  The details of our numerical simulations are presented in Appendix~\ref{app:simulation}. After the application of a large cut on missing transverse energy, we find the leading backgrounds are $\overline{t}t+{\rm jets}$ (one leptonic top and one hadronic top) and $\overline{t}t+Z$ (two hadronic tops, $Z$ decays to two neutrinos). The basic cuts to select events impose the following requirements:
\begin{itemize}
 \setlength\itemsep{0.5mm}
  \item[(a)] Missing energy $\slashed{E}_{\rm T} > 400$ GeV.
  \item[(b)] Using our modified algorithm, require at least two top-tagged jets with $p_T(J) > 200$~GeV.
  \item[(c)] At least two anti-$k_{\rm T}$, $R = 0.4$ $b$-tagged jets with $p_T$ above 30 GeV.
  \item[(d)] Veto events with isolated leptons with $p_{T, \ell} > 15$~GeV and $|\eta_{\ell}| < 2.4$.
\end{itemize}
For simplicity, we have implemented a $b$-tagging efficiency of 0.8 (mistag efficiency of 0.2 for charm quarks and 0.05 for light quarks and gluon)~\cite{CMS-PAS-BTV-15-001,ATLAS-CONF-2016-077} in our simulation. One could also perform direct $b$-tagging on the subjets of the fat jets as in Ref.~\cite{CMS:2014fya}. Here, the main requirement of our simplistic and conservative $b$-tagging approach is to sufficiently reduce the $W/Z+\mbox{jets}$ backgrounds by requiring two $b$-tags in the event. 

\begin{figure}[ht!]
\begin{center}
\includegraphics[scale=0.45]{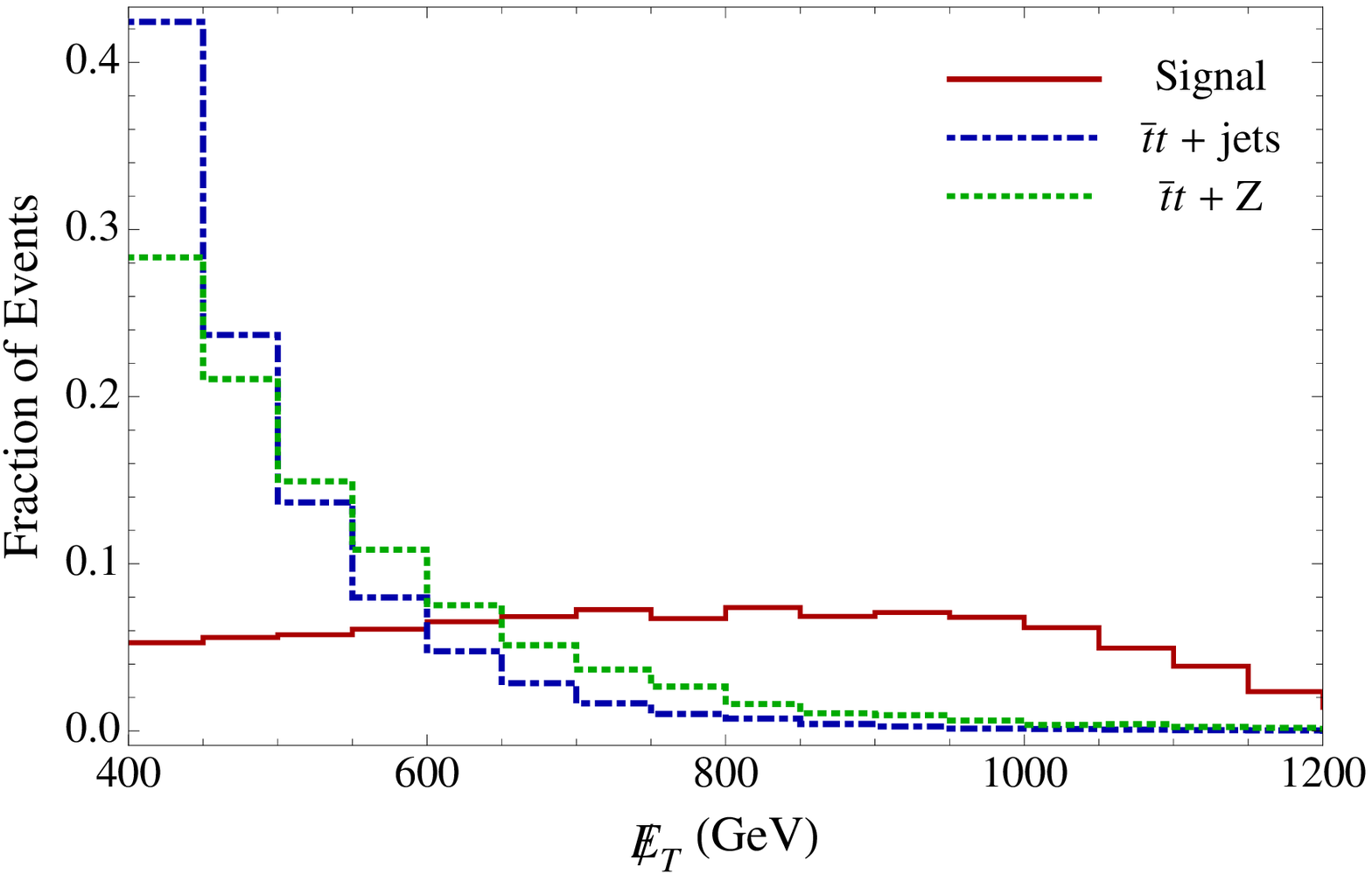} \hspace{5mm}
\includegraphics[scale=0.45]{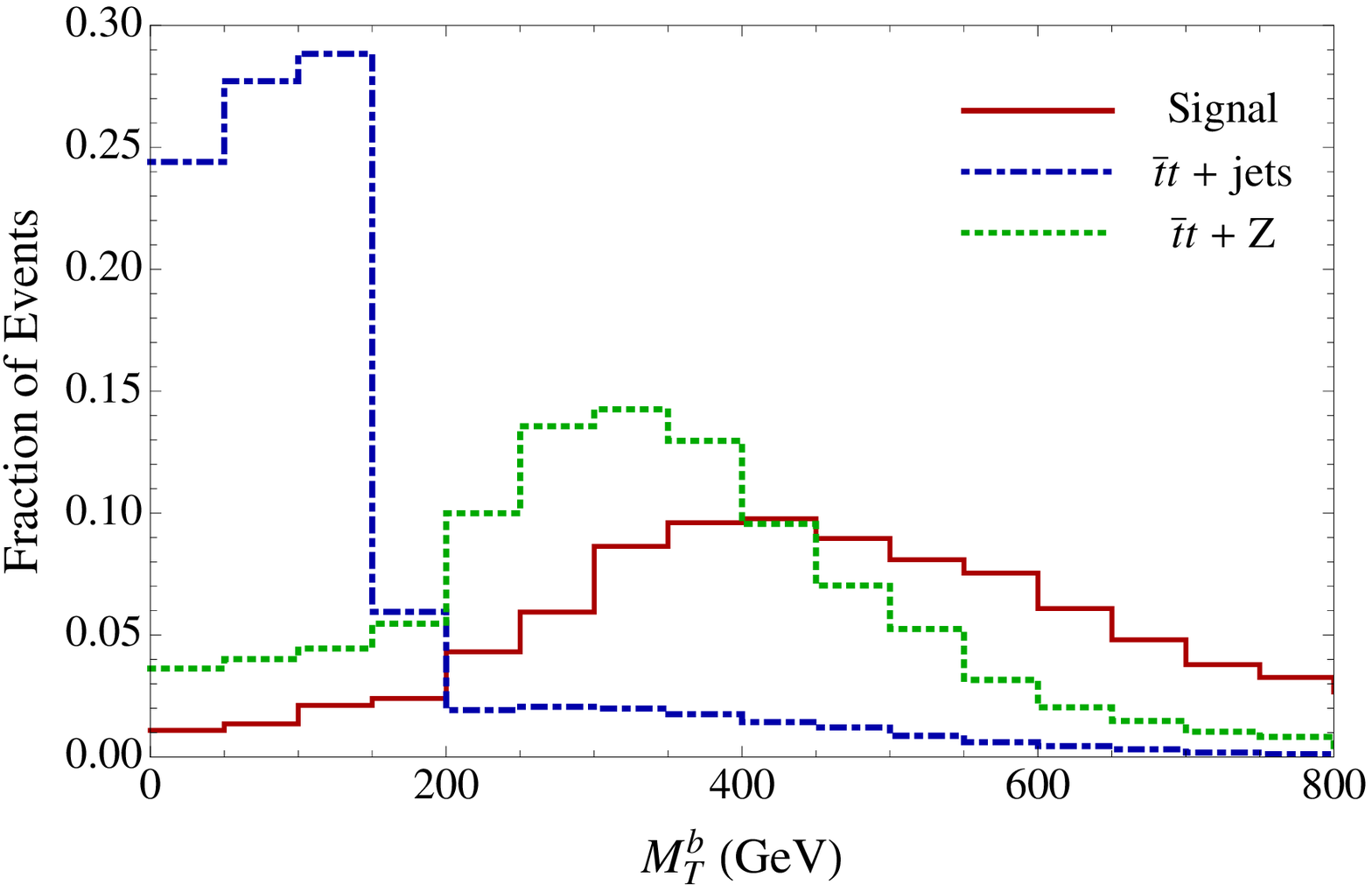}
\end{center}
\caption{Left panel: The normalized event distributions in $\slashed{E}_{\rm T}$ for the signal and leading backgrounds after basic cuts. Right panel: Same as the left but in terms of $M_T^b$. The signal has $m_{\widetilde{t}} = 1.2$~TeV and $m_{\widetilde{\chi}}=100$~GeV.}
\label{fig:METandMTb}
\end{figure}

After the basic cuts, we note two powerful kinematical variables which are useful for reducing backgrounds. The most obvious one is $\slashed{E}_{\rm T}$, which is shown in the left panel of Fig.~\ref{fig:METandMTb}. The other variable is the transverse mass associated with the $b$-jets and $\slashed{E}_{\rm T}$, which has already been adopted by both CMS~\cite{CMS-PAS-SUS-16-029} and ATLAS~\cite{ATLAS-CONF-2016-077}. It is given by
\beqa
M_T^b = \mbox{min}\big[ M_T(\vec{p}_{b_1}, \vec{\slashed{E}}_{\rm T} ),  M_T(\vec{p}_{b_2},  \vec{\slashed{E}}_{\rm T} ) \big] \,.
\eeqa
The transverse mass is defined as $M_T(\vec{p}_{b_i}, \vec{\slashed{E}}_{\rm T} ) = 2 p_T^{b_i} \, \slashed{E}_{\rm T}[ 1- \cos{(\phi_{b_i} - \phi_{\slashed{E}_{\rm T}}) }]$, where $\phi_{b_i}$ and $\phi_{\slashed{E}_{\rm T}}$ are the azimuthal angles of the $b$-jet and $\vec{\slashed{E}}_{\rm T}$. In the right-panel of Fig.~\ref{fig:METandMTb}, we show the signal and background distributions for this variable. 
As one can see, the $\overline{t}t + {\rm jets}$ background has an endpoint around the top quark mass, so requiring a sufficiently large $M_T^b$ can dramatically reduce this background. We also note that the tail of the  $M_T^b$ distribution for $\overline{t}t + {\rm jets}$ comes from mis-tagged $b$-jets, which do not come from top quark decays and therefore are not subject to this kinematic cutoff. 

\begin{figure}[ht!]
\begin{center}
\includegraphics[scale=0.6]{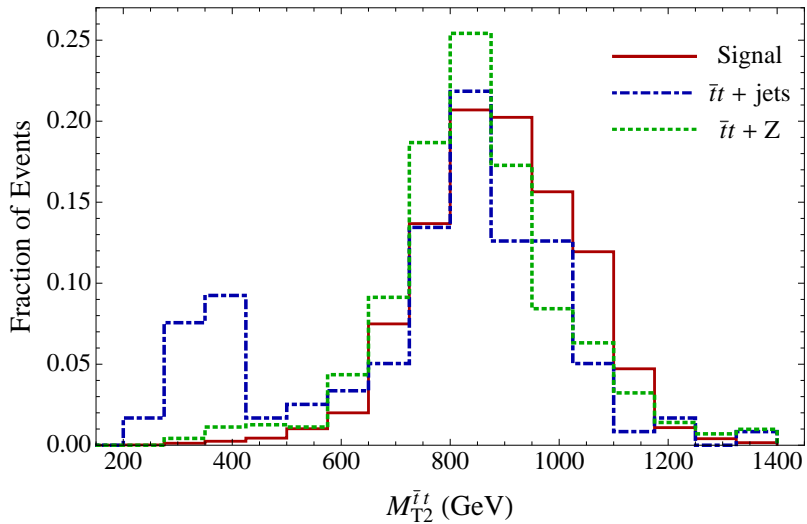}
\end{center}
\caption{Distributions for $M_{T2}^{\overline{t} t}$ after the optimized cuts of $\slashed{E}_{\rm T} > 800$~GeV and $M_{T}^{b} > 240$~GeV for a 1.2 TeV stop mass.}
\label{fig:mt2}
\end{figure}

With our new tagging algorithm, one can define additional kinematic variables based on the two top quarks in the final state. With the large missing transverse energy at hand, one natural choice is the $M_{T2}$-like variable. Specifically for events with two tops and missing transverse energy, one can define the following $M_{T2}^{\overline{t} t}$ variable as~\cite{Lester:1999tx,Barr:2003rg,Cheng:2008hk}
\beqa
M^{\overline{t} t }_{T2} = \mbox{min}\left\{   \bigcup_{ \vec{p}^T_1 + \vec{p}^T_2 = \vec{\slashed{E}}_{\rm T}   } \mbox{max} {\Big[} M_T(\vec{p}_{t_1},  \vec{p}^T_1), M_T(\vec{p}_{t_2},  \vec{p}^T_2) {\Big]}  \right\}  \, .
\label{eq:mt2}
\eeqa
In Fig.~\ref{fig:mt2}, we show the distribution for $M_{T2}^{\overline{t} t}$ for both the signal and leading backgrounds. The signal event distribution has a peak feature with an end-point of around 1.2 TeV, which can be easily understood from the stop mass and the large missing energy cut of $\slashed{E}_{\rm T} > 800$~GeV. The $\overline{t}t+Z$ background  has a similar distribution to the signal events and behaves as an irreducible background. For the $\overline{t}t + {\rm jets}$ background, two peak structures appear in the distribution. The peak at higher values is similar to the signal region of phase space with two top-jets moving in the same direction, while the peak at lower values is due to the region of phase space where one top-jet approximately aligns with the missing transverse momentum, leading to a smaller transverse mass.
Imposing a lower limit cut on $M_{T2}^{\overline{t} t}$ can therefore reduce the $\overline{t}t + {\rm jets}$ background. 

Using the kinematic variables outlined above, we perform an optimization of cuts to increase the expected signal discovery significance.  Since we expect small numbers of signal and background events, we use the following expected log likelihood discovery significance
\beqa
\sigma_{\rm LL} \equiv \sqrt{ - 2 \left[ S + (B + S) \log{\frac{B}{B + S}} \right]} \,,
\label{eq:sigmaLL}
\eeqa
which approaches $S/\sqrt{B}$ for large signal and background.   After optimizing the cuts on the three variables $\slashed{E}_{\rm T}$, $M_{T}^{b}$ and $M_{T2}^{\overline{t} t}$, we show the cut-flow of events in Table~\ref{tab:cut-flow-our}, which shows that requiring two top-jets from our modified tagging algorithm along with the rest of the basic cuts already reduces the dominant $\overline{t} t + {\rm jets}$ background substantially.
\begin{table}[h!]
\renewcommand{\arraystretch}{1.3}
\begin{center}
    \resizebox{\linewidth}{!}{
  \begin{tabular}{|  c | c | c| c | c || c |}
  \hline
Modified top-tagging  &  after basic cuts & $\slashed{E}_{\rm T} > 800$ GeV & $M_{T}^{b} > 240$ GeV & $M_{T2}^{\overline{t} t}> 500 $ GeV &  events ($100 \text{ fb}^{-1}$)  \\
    \hline     \hline
    signal &  0.17 fb & 0.084 fb & 0.081 fb & 0.080 fb & 8.0 \\
    \hline \hline
    $\overline{t} \, t \, +{\rm jets}$ & 4.66 fb & 0.090 fb & $9.8\times 10^{-3}$ fb & $7.8\times 10^{-3}$ fb & 0.78  \\
    \hline
    $\overline{t} \, t \, +Z$  & 0.17 fb & 0.010 fb & $9.1\times 10^{-3}$ fb & $8.8\times 10^{-3}$ fb & 0.88  \\
    \hline
    $Z+\mbox{jets}$ &  0.024 fb &  $1.3\times 10^{-3}$  fb & $1.0\times 10^{-3}$ fb & $1.0\times 10^{-3}$ fb & 0.1  \\
    \hline
    $W+\mbox{jets}$ &  $6.4\times 10^{-3}$ fb &  $0.2\times 10^{-3}$ fb & $0.2\times 10^{-3}$  fb & $0.2\times 10^{-3}$ fb & 0.02  \\
    \hline
    \hline
   \multicolumn{5}{|c|}{} &     \multicolumn{1}{|c|}{$S/B$ = 4.5} \\ 
      \multicolumn{5}{|c|}{} &     \multicolumn{1}{|c|}{$\sigma_{\rm LL}$ = 4.2}  \\
     \hline
  \end{tabular}
  }
  \end{center}
    \caption{Signal and background cross sections after cuts using our modified algorithm at the $\sqrt{s} = 13$ TeV LHC. The choice of cuts is optimized to increase $\sigma_{\rm LL}$ in Eq.~(\ref{eq:sigmaLL}). Here, we have $m_{\widetilde{t}} = 1.2$~TeV and $m_{\widetilde{\chi}}=100$~GeV.}
  \label{tab:cut-flow-our}
  \end{table}
The cuts on $\slashed{E}_{\rm T}$ and $M_{T}^{b}$  are still the most efficient for increasing $S/B$. As a result, large values of $S/B$ and $\sigma_{\rm LL}$ are obtained. The chosen model parameter with $m_{\widetilde{t}} = 1.2$~TeV and $m_{\widetilde{\chi}}=100$~GeV will be tested at the LHC Run 2 with 100~fb$^{-1}$. 
 
To compare to a search using the original HEPTopTagger algorithm, we replace part (b) of the basic cuts by
\begin{itemize}
\item[(b$'$)] Require at least one top-tagged jet with $p_T(J) > 200$~GeV based on the original HEPTopTagger algorithm,
 \end{itemize}
  and show the cut-flow of signal and backgrounds in Table~\ref{tab:cut-flow-original}. Comparing the results from Table~\ref{tab:cut-flow-our} and Table~\ref{tab:cut-flow-original}, one can see that our modified top-tagging algorithm yields an obvious improvement for $S/B$ and a mild increase for $\sigma_{\rm LL}$.  
\begin{table}[h!]
\renewcommand{\arraystretch}{1.3}
\begin{center}
    \resizebox{0.8\linewidth}{!}{
  \begin{tabular}{|  c | c | c| c || c |}
  \hline
Original top-tagging  &  after basic cuts & $\slashed{E}_{\rm T} > 800$ GeV & $M_{T}^{b} > 260$ GeV  &  events ($100 \text{ fb}^{-1}$)  \\
    \hline     \hline
    signal &  0.24 fb & 0.11 fb & 0.10 fb &  10.0 \\
    \hline \hline
    $\overline{t} \, t \, +{\rm jets}$ & 13.6 fb & 0.19 fb & $0.0174$ fb &  1.74  \\
    \hline
    $\overline{t} \, t \, +Z$  & 0.443 fb & 0.016 fb & 0.014 fb &  1.35  \\
    \hline
    $Z+\mbox{jets}$ &  0.164 fb &  $3.2\times 10^{-3}$  fb & $1.6\times 10^{-3}$ fb &  0.16  \\
    \hline
    $W+\mbox{jets}$ &  0.047 fb &  $1.2\times 10^{-3}$ fb & $0.6\times 10^{-3}$  fb &  0.06  \\
    \hline
    \hline
   \multicolumn{4}{|c|}{} &     \multicolumn{1}{|c|}{$S/B$ = 3.0} \\ 
      \multicolumn{4}{|c|}{} &     \multicolumn{1}{|c|}{$\sigma_{\rm LL}$ = 4.1}  \\
     \hline
  \end{tabular}
  }
  \end{center}
    \caption{The same as Table~\ref{tab:cut-flow-our} but based on the original HEPTopTagger algorithm.}
  \label{tab:cut-flow-original}
  \end{table}
  %

\section{Discussion and Conclusions}
\label{sec:conc}
We note that the signal acceptance obtained when requiring two top tagged jets is not that high compared to the original HEPTopTagger algorithm with the requirement of only one top tagged jet. This is simply due to the fact that a non-negligible fraction of top quarks from 1.2 TeV stop decays are not boosted. These events can have the partons from top decays well separated from each other. For instance, the $b$-hadron could be well separated from the hadronic $W$ and may not have enough transverse momentum to satisfy the fat jet $p_T > 200$~GeV requirement. In principle, one could include those events with an ordinary $b$-jet that does not belong to any fat jet. With a simple implementation, we have found the signal acceptance can be further increased by around 13\% with a slightly smaller value of $S/B=4.4$, but slightly larger value of $\sigma_{\rm LL} = 4.4$. If one wants to further increase the discovery sensitivity, one should combine events with boosted and non-boosted top quarks~\cite{CMS-PAS-SUS-16-029}.  

In addition, our algorithm should be validated with a full detector simulation before use.  The simplified approach to detector effects, as well as to $b$-tagging, adopted in this work is justified by the fact that the dominant smearing of jet kinematics is coming from QCD physics, but there is some effect of calorimetric smearing as well.  Our work demonstrates the gains to be had by adopting our algorithm even neglecting these effects and knowing that the final projections for discovery sensitivity are an estimation.

In this paper, we have concentrated on simplified models with stops and neutralinos. However, the key observation illustrated in Fig.~\ref{fig:schematic} is general and could be applied to other models, under the condition that a sizeable cut on the missing transverse energy prefers to have the visible particles collimated. One possible example is pair-produced gluinos with off-shell squark mediated decay, $\widetilde{g}\rightarrow jj\widetilde{\chi}_{0}$. The four jets in the final state could merge into a single fat jet after a sufficiently large cut on $\slashed{E}_{\rm T}$. A similar jet-substructure based analysis is therefore ideal to improve the signal discovery significance.  

In summary, we have identified a new and interesting region of phase space for the heavy stop plus light neutralino model. The subset of signal events with large missing transverse momentum has a region of phase space where the partons coming from the decay of the two hadronic top quarks have a large overlapping. We have developed a jet-substructure based algorithm to identify these two merged or semi-merged hadronic tops. Based on our estimation of signal discovery significance with only statistical errors, we have found that our algorithm can dramatically increase $S/B$ by 50\% and yield a mild increase in the discovery significance. It is also important to point out that our new algorithm is ideal for use with additional kinematic variables which require the complete reconstruction of the two hadronic top momenta to further increase $S/B$ and the discovery significance.  A stop mass of 1.2 TeV and a light neutralino will be  concretely tested at the LHC Run 2 with 100 fb$^{-1}$.

\subsection*{Acknowledgments}
We would like to thank Vernon Barger and Tilman Plehn for discussion and comments. This work is supported by the U. S. Department of Energy under the contract DE-FG-02-95ER40896. This work was partially performed at the Aspen Center for Physics, which is supported by National Science Foundation grant PHY-1066293.

\appendix
\section{Simulation Details}
\label{app:simulation}
We use \texttt{MadGraph5\_aMC@NLO} \cite{Alwall:2014hca} package to simulate all parton level events. Showering of the parton level events was done in \texttt{Pythia8} \cite{Sjostrand:2007gs}. After showering, a $0.1\times 0.1$ detector granularization was applied for particles in the final state, which were then analyzed and clustered using \texttt{FastJet} \cite{Cacciari:2011ma}. For the signal, we simulated 50000 events at a center-of-mass energy of 13 TeV. We assume a $K$-factor of 1.5 for the signal production cross section~\cite{Borschensky:2014cia}. 
The desired signal topology where the stops decay with a branching ratio of 100\% to tops and neutralinos, with the tops decaying hadronically, was forced in the simulation. All other parameters in the signal simulation were left at their default values. 
 
 For all background events, we simulate events at a center-of-mass energy of 13 TeV, using the default SM model file, with default model values. For the $\overline{t} \, t \, + \ts{\rm jets}$ background, we simulated $4.975\times 10^7$ events. We used an unmatched sample involving 2 jets, which we checked the variable distributions for a matched sample after the final cuts. We assume a $K$-factor of 1.3 \cite{Alwall:2014hca}. The desired semi-leptonic background topology where one top decays leptonically (without taus)  and the other hadronically was forced in the simulation. All other parameters in the $\overline{t} \, t \, + {\rm jets}$ simulation were left at their default values, except for imposing a generator level cut requiring the missing transverse energy (sum of the neutrino momenta) to be above 200 GeV.  For the $\overline{t} \, t \, +Z$ background, we simulated $7.5\times 10^5$ events. The desired background topology where both tops decay hadronically and the $Z$ boson decays to neutrinos was forced in the simulation. All other parameters in the $\overline{t} \, t \, +Z$ simulation were left at their default values, except requiring a baseline cut of $\slashed{E}_{\rm T}>200$~GeV.  For the $Z+\ts{\rm jets}$ background, we simulated $3.75\times 10^7$ events. The desired background topology where the $Z$ boson decays to neutrinos was forced in the simulation. All other parameters in the $Z+3j$ simulation were left at their default values, except requiring $\slashed{E}_{\rm T}>200$~GeV. For the $W+\ts {\rm jets}$ background, we simulated $4.975\times 10^7$ events. The desired background topology where the $W$ boson decays to $\ell \bar{\nu}$ or $\bar{\ell} \nu$, where $\ell$ does not include taus, was forced in the simulation. All other parameters in the $W+3j$ simulation were left at their default values, except a cut of $\slashed{E}_{\rm T}>200$~GeV.


\begin{thebibliography}{10}

\bibitem{CMS:2012nga}
{\bf CMS} Collaboration, S.~Chatrchyan et~al., {\it {A New Boson with a Mass of
  125 GeV Observed with the CMS Experiment at the Large Hadron Collider}},
  {\em Science} {\bf 338} (2012) 1569--1575.

\bibitem{ATLAS:2012oga}
{\bf ATLAS} Collaboration, G.~Aad et~al., {\it {A particle consistent with the
  Higgs Boson observed with the ATLAS Detector at the Large Hadron Collider}},
  {\em Science} {\bf 338} (2012) 1576--1582.

\bibitem{Martin:1997ns}
S.~P. Martin, {\it {A Supersymmetry primer}},
  \href{http://arxiv.org/abs/hep-ph/9709356}{{\tt hep-ph/9709356}}. [Adv. Ser.
  Direct. High Energy Phys.18,1(1998)].

\bibitem{Chung:2003fi}
D.~J.~H. Chung, L.~L. Everett, G.~L. Kane, S.~F. King, J.~D. Lykken, and L.-T.
  Wang, {\it {The Soft supersymmetry breaking Lagrangian: Theory and
  applications}},  {\em Phys. Rept.} {\bf 407} (2005) 1--203,
  [\href{http://arxiv.org/abs/hep-ph/0312378}{{\tt hep-ph/0312378}}].

\bibitem{Dimopoulos:1995mi}
S.~Dimopoulos and G.~F. Giudice, {\it {Naturalness constraints in
  supersymmetric theories with nonuniversal soft terms}},  {\em Phys. Lett.}
  {\bf B357} (1995) 573--578, [\href{http://arxiv.org/abs/hep-ph/9507282}{{\tt
  hep-ph/9507282}}].

\bibitem{Cohen:1996vb}
A.~G. Cohen, D.~B. Kaplan, and A.~E. Nelson, {\it {The More minimal
  supersymmetric standard model}},  {\em Phys. Lett.} {\bf B388} (1996)
  588--598, [\href{http://arxiv.org/abs/hep-ph/9607394}{{\tt hep-ph/9607394}}].

\bibitem{Cahill-Rowley:2014twa}
M.~Cahill-Rowley, J.~L. Hewett, A.~Ismail, and T.~G. Rizzo, {\it {Lessons and
  prospects from the pMSSM after LHC Run I}},  {\em Phys. Rev.} {\bf D91}
  (2015), no.~5 055002, [\href{http://arxiv.org/abs/1407.4130}{{\tt
  arXiv:1407.4130}}].

\bibitem{Baer:2014ica}
H.~Baer, V.~Barger, D.~Mickelson, and M.~Padeffke-Kirkland, {\it {SUSY models
  under siege: LHC constraints and electroweak fine-tuning}},  {\em Phys. Rev.}
  {\bf D89} (2014), no.~11 115019, [\href{http://arxiv.org/abs/1404.2277}{{\tt
  arXiv:1404.2277}}].

\bibitem{Han:2012fw}
Z.~Han, A.~Katz, D.~Krohn, and M.~Reece, {\it {(Light) Stop Signs}},  {\em
  JHEP} {\bf 08} (2012) 083, [\href{http://arxiv.org/abs/1205.5808}{{\tt
  arXiv:1205.5808}}].

\bibitem{Delgado:2012eu}
A.~Delgado, G.~F. Giudice, G.~Isidori, M.~Pierini, and A.~Strumia, {\it {The
  light stop window}},  {\em Eur. Phys. J.} {\bf C73} (2013), no.~3 2370,
  [\href{http://arxiv.org/abs/1212.6847}{{\tt arXiv:1212.6847}}].

\bibitem{Dutta:2013gga}
B.~Dutta, W.~Flanagan, A.~Gurrola, W.~Johns, T.~Kamon, P.~Sheldon, K.~Sinha,
  K.~Wang, and S.~Wu, {\it {Probing compressed top squark scenarios at the LHC
  at 14 TeV}},  {\em Phys. Rev.} {\bf D90} (2014), no.~9 095022,
  [\href{http://arxiv.org/abs/1312.1348}{{\tt arXiv:1312.1348}}].

\bibitem{Cho:2014yma}
W.~S. Cho, J.~S. Gainer, D.~Kim, K.~T. Matchev, F.~Moortgat, L.~Pape, and
  M.~Park, {\it {Improving the sensitivity of stop searches with on-shell
  constrained invariant mass variables}},  {\em JHEP} {\bf 05} (2015) 040,
  [\href{http://arxiv.org/abs/1411.0664}{{\tt arXiv:1411.0664}}].

\bibitem{Ferretti:2015dea}
G.~Ferretti, R.~Franceschini, C.~Petersson, and R.~Torre, {\it {Spot the stop
  with a b-tag}},  {\em Phys. Rev. Lett.} {\bf 114} (2015) 201801,
  [\href{http://arxiv.org/abs/1502.01721}{{\tt arXiv:1502.01721}}].

\bibitem{An:2015uwa}
H.~An and L.-T. Wang, {\it {Opening up the compressed region of top squark
  searches at 13 TeV LHC}},  {\em Phys. Rev. Lett.} {\bf 115} (2015) 181602,
  [\href{http://arxiv.org/abs/1506.00653}{{\tt arXiv:1506.00653}}].

\bibitem{Belanger:2015vwa}
G.~Belanger, D.~Ghosh, R.~Godbole, and S.~Kulkarni, {\it {Light stop in the
  MSSM after LHC Run 1}},  {\em JHEP} {\bf 09} (2015) 214,
  [\href{http://arxiv.org/abs/1506.00665}{{\tt arXiv:1506.00665}}].

\bibitem{Kobakhidze:2015scd}
A.~Kobakhidze, N.~Liu, L.~Wu, J.~M. Yang, and M.~Zhang, {\it {Closing up a
  light stop window in natural SUSY at LHC}},  {\em Phys. Lett.} {\bf B755}
  (2016) 76--81, [\href{http://arxiv.org/abs/1511.02371}{{\tt
  arXiv:1511.02371}}].

\bibitem{Cheng:2016npb}
H.-C. Cheng, L.~Li, and Q.~Qin, {\it {Second Stop and Sbottom Searches with a
  Stealth Stop}},  \href{http://arxiv.org/abs/1607.06547}{{\tt
  arXiv:1607.06547}}.

\bibitem{Plehn:2010st}
T.~Plehn, M.~Spannowsky, M.~Takeuchi, and D.~Zerwas, {\it {Stop Reconstruction
  with Tagged Tops}},  {\em JHEP} {\bf 10} (2010) 078,
  [\href{http://arxiv.org/abs/1006.2833}{{\tt arXiv:1006.2833}}].

\bibitem{Bi:2011ha}
X.-J. Bi, Q.-S. Yan, and P.-F. Yin, {\it {Probing Light Stop Pairs at the
  LHC}},  {\em Phys. Rev.} {\bf D85} (2012) 035005,
  [\href{http://arxiv.org/abs/1111.2250}{{\tt arXiv:1111.2250}}].

\bibitem{Plehn:2012pr}
T.~Plehn, M.~Spannowsky, and M.~Takeuchi, {\it {Stop searches in 2012}},  {\em
  JHEP} {\bf 08} (2012) 091, [\href{http://arxiv.org/abs/1205.2696}{{\tt
  arXiv:1205.2696}}].

\bibitem{Alves:2012ft}
D.~S.~M. Alves, M.~R. Buckley, P.~J. Fox, J.~D. Lykken, and C.-T. Yu, {\it
  {Stops and $\not E_T$: The shape of things to come}},  {\em Phys. Rev.} {\bf
  D87} (2013), no.~3 035016, [\href{http://arxiv.org/abs/1205.5805}{{\tt
  arXiv:1205.5805}}].

\bibitem{Kaplan:2012gd}
D.~E. Kaplan, K.~Rehermann, and D.~Stolarski, {\it {Searching for Direct Stop
  Production in Hadronic Top Data at the LHC}},  {\em JHEP} {\bf 07} (2012)
  119, [\href{http://arxiv.org/abs/1205.5816}{{\tt arXiv:1205.5816}}].

\bibitem{Buckley:2013lpa}
M.~R. Buckley, T.~Plehn, and M.~Takeuchi, {\it {Buckets of Tops}},  {\em JHEP}
  {\bf 08} (2013) 086, [\href{http://arxiv.org/abs/1302.6238}{{\tt
  arXiv:1302.6238}}].

\bibitem{ATLAS-CONF-2016-077}
{\bf ATLAS} Collaboration, {\it {Search for the Supersymmetric Partner of the
  Top Quark in the Jets+Emiss Final State at sqrt(s) = 13 TeV}},  Tech. Rep.
  ATLAS-CONF-2016-077, CERN, Geneva, Aug, 2016.

\bibitem{CMS-PAS-SUS-16-029}
{\bf CMS} Collaboration, {\it {Search for direct top squark pair production in
  the fully hadronic final state in proton-proton collisions at sqrt(s) = 13
  TeV corresponding to an integrated luminosity of 12.9/fb}},  Tech. Rep.
  CMS-PAS-SUS-16-029, CERN, Geneva, 2016.

\bibitem{Kasieczka:2015jma}
G.~Kasieczka, T.~Plehn, T.~Schell, T.~Strebler, and G.~P. Salam, {\it
  {Resonance Searches with an Updated Top Tagger}},  {\em JHEP} {\bf 06} (2015)
  203, [\href{http://arxiv.org/abs/1503.05921}{{\tt arXiv:1503.05921}}].

\bibitem{Lester:1999tx}
C.~G. Lester and D.~J. Summers, {\it {Measuring masses of semiinvisibly
  decaying particles pair produced at hadron colliders}},  {\em Phys. Lett.}
  {\bf B463} (1999) 99--103, [\href{http://arxiv.org/abs/hep-ph/9906349}{{\tt
  hep-ph/9906349}}].

\bibitem{Low:2013aza}
I.~Low, {\it {Polarized charginos (and top quarks) in scalar top quark
  decays}},  {\em Phys. Rev.} {\bf D88} (2013), no.~9 095018,
  [\href{http://arxiv.org/abs/1304.0491}{{\tt arXiv:1304.0491}}].

\bibitem{Dokshitzer:1997in}
Y.~L. Dokshitzer, G.~D. Leder, S.~Moretti, and B.~R. Webber, {\it {Better jet
  clustering algorithms}},  {\em JHEP} {\bf 08} (1997) 001,
  [\href{http://arxiv.org/abs/hep-ph/9707323}{{\tt hep-ph/9707323}}].

\bibitem{Thaler:2011gf}
J.~Thaler and K.~Van~Tilburg, {\it {Maximizing Boosted Top Identification by
  Minimizing N-subjettiness}},  {\em JHEP} {\bf 02} (2012) 093,
  [\href{http://arxiv.org/abs/1108.2701}{{\tt arXiv:1108.2701}}].

\bibitem{Butterworth:2008iy}
J.~M. Butterworth, A.~R. Davison, M.~Rubin, and G.~P. Salam, {\it {Jet
  substructure as a new Higgs search channel at the LHC}},  {\em Phys. Rev.
  Lett.} {\bf 100} (2008) 242001, [\href{http://arxiv.org/abs/0802.2470}{{\tt
  arXiv:0802.2470}}].

\bibitem{CMS:2014fya}
{\bf CMS} Collaboration, C.~Collaboration, {\it {Boosted Top Jet Tagging at
  CMS}}, .

\bibitem{CMS-PAS-BTV-15-001}
{\bf CMS} Collaboration, {\it {Identification of b quark jets at the CMS
  Experiment in the LHC Run 2}},  Tech. Rep. CMS-PAS-BTV-15-001, CERN, Geneva,
  2016.

\bibitem{Barr:2003rg}
A.~Barr, C.~Lester, and P.~Stephens, {\it {m(T2): The Truth behind the
  glamour}},  {\em J. Phys.} {\bf G29} (2003) 2343--2363,
  [\href{http://arxiv.org/abs/hep-ph/0304226}{{\tt hep-ph/0304226}}].

\bibitem{Cheng:2008hk}
H.-C. Cheng and Z.~Han, {\it {Minimal Kinematic Constraints and m(T2)}},  {\em
  JHEP} {\bf 12} (2008) 063, [\href{http://arxiv.org/abs/0810.5178}{{\tt
  arXiv:0810.5178}}].

\bibitem{Alwall:2014hca}
J.~Alwall, R.~Frederix, S.~Frixione, V.~Hirschi, F.~Maltoni, O.~Mattelaer,
  H.~S. Shao, T.~Stelzer, P.~Torrielli, and M.~Zaro, {\it {The automated
  computation of tree-level and next-to-leading order differential cross
  sections, and their matching to parton shower simulations}},  {\em JHEP} {\bf
  07} (2014) 079, [\href{http://arxiv.org/abs/1405.0301}{{\tt
  arXiv:1405.0301}}].

\bibitem{Sjostrand:2007gs}
T.~Sjostrand, S.~Mrenna, and P.~Z. Skands, {\it {A Brief Introduction to PYTHIA
  8.1}},  {\em JHEP05 (2006) 026, Comput. Phys. Commun.} {\bf 178} (2008)
  852--867, [\href{http://arxiv.org/abs/0710.3820}{{\tt arXiv:0710.3820}}].

\bibitem{Cacciari:2011ma}
M.~Cacciari, G.~P. Salam, and G.~Soyez, {\it {FastJet User Manual}},  {\em Eur.
  Phys. J.} {\bf C72} (2012) 1896, [\href{http://arxiv.org/abs/1111.6097}{{\tt
  arXiv:1111.6097}}].

\bibitem{Borschensky:2014cia}
C.~Borschensky, M.~Krämer, A.~Kulesza, M.~Mangano, S.~Padhi, T.~Plehn, and
  X.~Portell, {\it {Squark and gluino production cross sections in pp
  collisions at $\sqrt{s}$ = 13, 14, 33 and 100 TeV}},  {\em Eur. Phys. J.}
  {\bf C74} (2014), no.~12 3174, [\href{http://arxiv.org/abs/1407.5066}{{\tt
  arXiv:1407.5066}}].

\end{thebibliography}
\providecommand{\href}[2]{#2}\begingroup\raggedright\endgroup

 \end{document}